\documentclass[aps,reprint,showpacs,superscriptaddress,
footinbib,citeautoscript]{revtex4-1}

\usepackage{graphicx}

\usepackage[utf8]{inputenc}
\usepackage{csquotes}
\usepackage[american]{babel}
\usepackage[T1]{fontenc}
\usepackage{enumerate}
\usepackage{mdwlist}
\usepackage[activate=normal]{pdfcprot}
\usepackage{bbding}
\usepackage{color}
\usepackage{amssymb}
\usepackage{amsmath}
\usepackage{amsfonts}
\usepackage{mathrsfs}

\usepackage{bm}
\usepackage{dcolumn}
\usepackage{color}
\usepackage[colorlinks=true,citecolor=blue]{hyperref}
\hypersetup{colorlinks=true,citecolor=blue,linkcolor=red
,urlcolor=blue}

\begin{document}

\title{Optical absorption properties of few-layer phosphorene}

\author{Zahra Torbatian}
\affiliation{School of Nano Science, Institute for Research in Fundamental Sciences (IPM), Tehran 19395-5531, Iran}
\author{Reza Asgari}
\email{asgari@ipm.ir}
\affiliation{School of Nano Science, Institute for Research in Fundamental Sciences (IPM), Tehran 19395-5531, Iran}
\affiliation{School of Physics, Institute for Research in Fundamental Sciences (IPM), Tehran 19395-5531, Iran}

\begin{abstract}

We investigate the optical absorption and transmission of few-layer phosphorene in the framework of {\it ab initio} density functional simulations and many-body perturbation theory at the level of random phase approximation. In bilayer phosphorene, the optical transition of the valence band to the conduction band appears along the armchair direction at about
0.72 eV, while it is absent along the zigzag direction. This phenomenon is consistent with experimental observations. The angle-resolved optical absorption in few-layer
phosphorene shows that it is transparent when illuminated by near grazing incidence of light. Also, there is a general trend of an increase in the absorption by increasing
 the number of layers.
Our results show that the bilayer phosphorene exhibits greater absorbance compared to that of bilayer graphene in the ultraviolet region.
Moreover, the maximal peak in the calculated absorption of bilayer MoS$_2$ is in the visible region, while bilayer graphene and phosphorene are transparent.
Besides, the collective electronic excitations of few-layer phosphorene are explored.
An optical mode (in-phase mode) that follows a low-energy $\sqrt q$ dependence for all structures, and another which is a damped acoustic mode (out-of-phase mode) with linear dispersion for multilayer phosphorene are obtained.
 The anisotropy of the band structure of few-layer phosphorene along the armchair and zigzag directions is manifested in the collective plasmon excitations.
\end{abstract}

\maketitle

\section{Introduction}\label{sec:intro}

Advanced two-dimensional (2D) crystalline layered materials have received
great attention during the past decade. The most well-known 2D crystalline materials are
graphene~\cite{zhang,asgari} and molybdenum disulfide~\cite{wang_2012}. The gapless nature of graphene and a low carrier mobility of transition metal dichalcogenides (TMDCs) have limited their potential applications in technology and electronic devices~\cite{wang_2015}. Phosphorene, one or a few layers of black phosphorous~\cite{Liu}, is another 2D semiconductor that has lately become the focus of investigations owing to its high carrier mobilities and a tunable band gap~\cite{qiao,peng}. High charge mobility on the order of $10^5$ cm$^2$/Vs has been observed in monolayer phosphorene at low temperatures~\cite{warsch}. Also, the narrow gap of phosphorene (between zero-gap graphene and large-gap TMDCs), makes it an ideal material for near and mid-infrared optoelectronics and  new types of plasmonic devices~\cite{xia2014}.

Similar to graphene and TMDCs, phosphorene is also a layered material that can be exfoliated to yield individual layers \cite{Lu}.
Monolayer phosphorene with a hexagonal puckered lattice has a direct band gap and high anisotropic band structure. The dispersion relation in monolayer phosphorene is highly anisotropic, which gives rise to the direction depending on  mechanical, optical, electronic and transport properties~\cite{wang_nano,jin_2016,zare_2017,nourbakhsh}. Bilayer phosphorene has attracted considerable interest given its potential application in nanoelectronics owing to its natural band gap and high carrier mobility~\cite{zhang_bi,Tran2014,Dai_bi}.
Bilayer phosphorene has a smaller direct band gap than monolayer phosphorene, which offers more opportunity to tune the semiconductor, metal and maybe even topological insulator~\cite{Liu}. More efforts have been devoted to the band structure and electronic properties of few-layer phosphorene, however, its optical properties deserve special consideration. Most importantly, phosphorene exhibits strong light-matter interactions in the visible and infrared photon energies. The application of few-layer phosphorene in nanoplasmonics and terahertz devices is highly promising~\cite{Buscema,Tran}.

The aim of this paper is to explore the angle-resolved optical absorption and transmission of few-layer phosphorene.
To do so, we use a recently proposed theoretical formulation \cite{Novko2016}.
In this theory, the current-current response tensor is calculated in the framework of {\it ab initio} density-functional theory (DFT) within a many-body
random-phase approximation (RPA), where the electromagnetic interaction is mediated by the free-photon propagator.
The tensorial character of the theory allows us to investigate the response to a transverse electric, $s$(TE), and transverse magnetic, $p$(TM), external electromagnetic field, separately. It is important to mention that the present theory doesn't include quasiparticle correction of the DFT band structure
such as an electron-hole bound state or excitons in semiconducting two-dimensional crystalline materials.
Also, it is possible to explore the dependence of the optical properties of the system on the incident angle of the external electromagnetic field.

For the sake of completeness, we calculate the collective modes of few-layer phosphorene. We use the formalism that has been presented in previous works \cite {Novko_2015,torbatian,torbatian2018}
and calculate the density-density response function within the DFT-RPA approach. Eventually, invoking the density-density response function, the collective modes are established by the zero of the real part of the macroscopic dielectric function.

In this work,  the optical absorption and transmission in terms of the {\it ab initio} current-current response tensor are calculated for few-layer phosphorene.
We show that the optical absorbance of the systems monotonically decreases, as the incident angle of light increases and few-layer phosphorene is transparent when it is illuminated by near grazing incidence of light. But the transmission increases, as the incident angle of light and it becomes almost $100\%$ for near grazing incidence. In the low-energy zone, the absorption spectrum is red-shifted by increasing the number of layers
along the armchair direction, while it changes slightly along the zigzag direction.
In addition, the plasmon excitations show a highly anisotropic plasmon dispersion owing to the strong crystal anisotropy in few-layer phosphorene. Finally, we make a brief comparison of the plasmon dispersion and optical absorption in few-layer phosphorene.

This paper is organized as follows.
In Sec.~\ref{sec:model}A, we present the methodology used for calculating the ground-state properties of the system.  In Sec.~\ref{sec:model}B, we present the optical quantities such as the optical absorption and transmission which are needed for the study in this work.
In Sec.~\ref{sec:result} we present and discuss in details our numerical results for the optical absorption spectra for pristine few-layer phosphorene.
Last but not least, we wrap up our main results in Sec.~\ref{sec:concl}.

\section{THEORY AND COMPUTATIONAL METHODS}\label{sec:model}
\subsection{Structure and ground-stat calculations}

In order to investigate the structural and optical properties of few-layer phosphorene, we use {\it ab initio} simulations, based on the density functional theory as implemented in the QUANTUM ESPRESSO~\cite{qE} code. Our DFT calculations are carried out using the Perdew-Burke-Ernzerh  exchange-correlation functional~\cite{PBE} coupled with the DFT-van der Waals method. The Kohn-Sham orbitals are expanded in a plane wave basis set with a cutoff energy which is $50$ Ry and a Monkhorst-Pack $k$-point mesh of $60\times80\times1$ is used. The energy convergence criteria for the electronic and ionic iterations are set to be $10^{-5}$ and $10^{-4}$ eV, respectively. In order to minimize the interaction between layers, a vacuum space of $20$ \AA~is applied.

In single layer phosphorene, each phosphorus atom covalently bonds with three adjacent atoms, forming a $sp^3$ hybridized puckered honeycomb structure. Therefore, it contains two atomic layers and two kinds of bonds with $2.22$ and $2.24$\AA~ bond lengths for in-plane and inter-plane $p-p$ connections, respectively. In the case of monolayer phosphorene, the calculated lattice constants are found to be $a = 4.62$ and  $b = 3.29$\AA~ along the armchair and zigzag directions, respectively, which are in good agreement with those predicted by other theoretical work~\cite{Elahi}.

\begin{figure}
\includegraphics*[width=8.7cm]{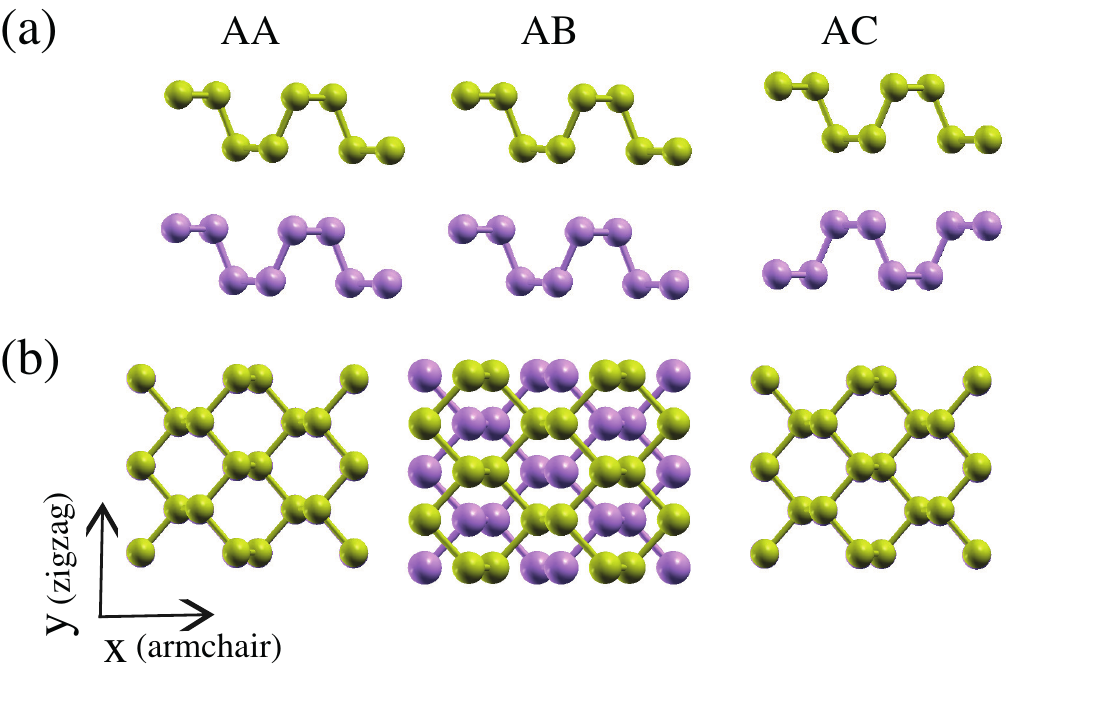}
\caption{(Color online) Three stacking structures of bilayer phosphorene. (a) Top view of AA, AB and AC stacking (b) Side view of AA, AB and AC stacking. Our numerical results of the ground-state energy show that the AB-stacked structure is energetically the most preferred for bilayer phosphorene.}\label{fig1}
\end{figure}

Bilayer phosphoren has been predicted to exist in three different stacking, namely AA, AB and AC structures (Fig.\ref{fig1}). For the AA stacking, the top layer is directly stacked on the bottom layer. The AB stacking can be viewed as shifting the bottom layer of the AA stacking by half of the cell along either the $a$ or $b$ directions. For the AC stacking, the top and bottom layers are mirror images of each other. Regarding the trilayer phosphorene, three possible stacking types (AAB, ABA, and ACA) are considered.

The ground-state energy calculations, stemming from our numerical calculations, show that the AB-stacked and ABA-stacked structures are energetically the most preferred for bilayer and trilayer phosphorene, respectively. For bilayer phosphorene with the AB-stacked structure, the interlayer separation between the closest phosphorene atoms is $3.55$\AA~ while the in-plane lattice constants remain the same. Throughout this study, we perform calculations on the most stable AB-stacked form of the bilayer and ABA-stacked form of trilayer phosphorene systems.

\begin{figure}
\includegraphics*[width=8.7cm]{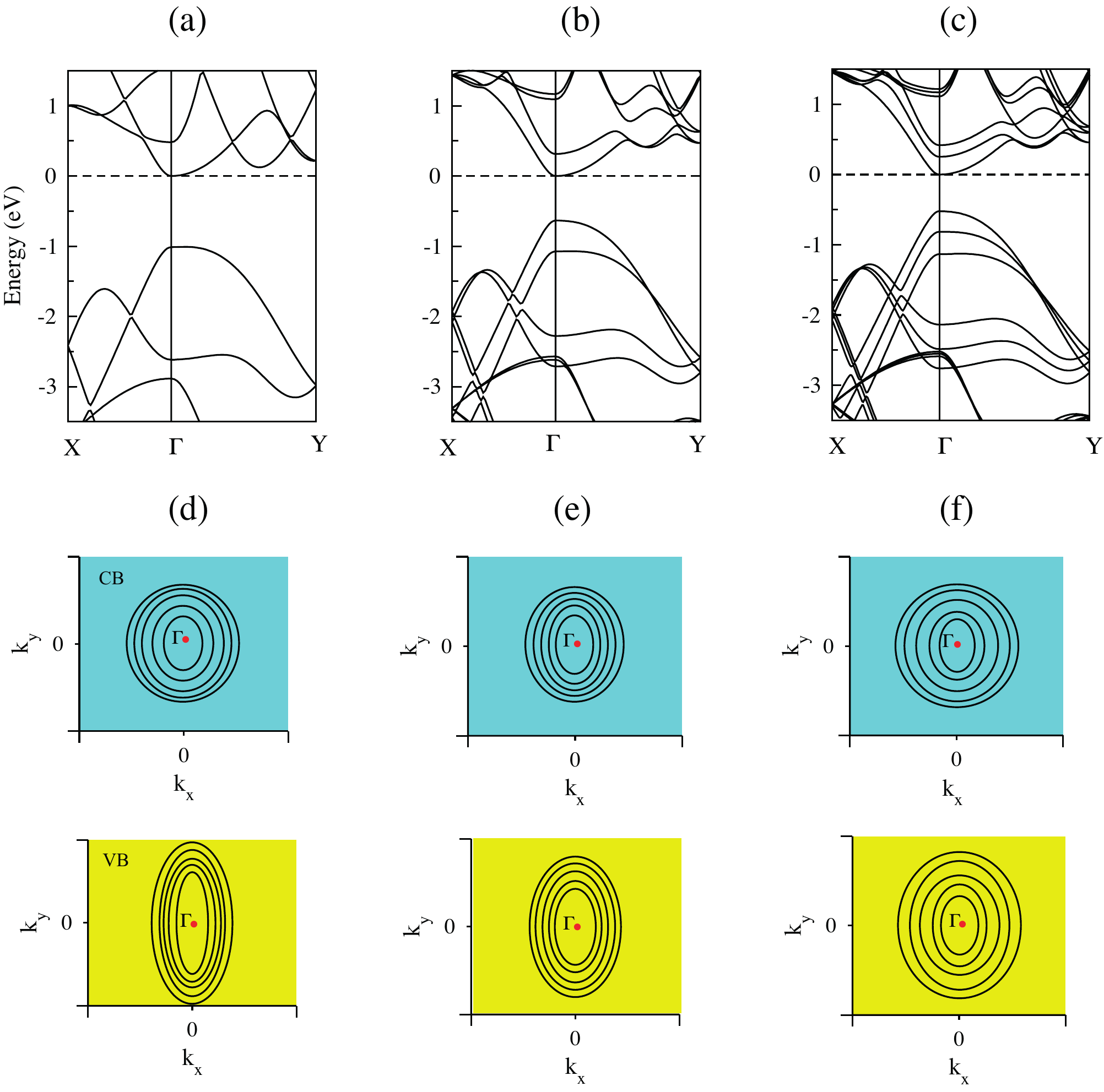}
\caption{(Color online) The calculated band structures for (a) monolayer (b) bilayer and (c) trilayer phosphorene structures along the $X-\Gamma-Y$ direction. The band gap for monolayer phosphorene is 0.98 eV and it decreases to 0.6 and 0.52 eV for bilayer and trilayer phosphorene, respectively. The nature of the band gap remains direct for trilayer to single-layer black phosphorous.
Isofrequency contour plots of the energies around the conduction band (CB) minimum and the valence band (VB) maximum in the k-space are illustrated for the (d) monolayer, (e) bilayer and (f) trilayer phosphorene for $E_{\rm F}= 0.11$ eV and $E_{\rm F}=-0.11$ eV, with a step of $0.02$ eV. The shape of the Fermi surfaces in the systems is almost an elliptic shape especially at low charge density.
}\label{fig2}
\end{figure}

\subsection{Current-current response tensor}
In this section, we consider independent electrons which live in a local crystal potential obtained by DFT and interact with the electromagnetic field described by the vector potential, $A({\bf r}, t)$. We pursue the procedure given in \cite{Novko2016} and thus the screened current-current response tensor can be calculated by solving the Dyson equation

\begin{eqnarray}
\Pi=\Pi_0 + \Pi_0 \otimes \mathbf D_0 \otimes \Pi
\label{dyson}
\end{eqnarray}
where $\Pi_0$ and $\mathbf D_0$ are the noninteracting current-current response tensor and free-photon propagator, respectively.
Our system contains slabs repeated periodically and $\Pi_{\mu\nu}$ is a periodic function along the $z$ direction. Also, the photon propagator has a long-range character that leads to the interactions between adjacent crystal slabs. Therefore, the screened $\Pi_{\mu\nu}$ contains the effects of the coupling between supercells. One of the best idea to solve this problem is to suppose that supercells are not repeated periodically along the $z$ direction and the system has just one crystal slab, which is restricted to the region $-L/2 < z < L/2$. It denotes $\Pi_{\mu\nu}$ contains just one term in the $z$ direction and photon propagator $D^0_{\mu\nu}$ couples only to the charge or current fluctuations in the region $-L/2 < z < L/2$, even if it propagates interaction all over the space.

With this well-known method for 2D materials, although the periodicity is broken along the $z$ direction, it remains in the $x-y$ plane. Therefore, we can perform the Fourier transform of the Dyson equation (Eq. \ref{dyson}) in the $x-y$ plane and results are given by

\begin{eqnarray}
\Pi_{G_\parallel, G'_{\parallel}}(\mathbf q, \omega, z, z')= \Pi^0_{G_\parallel, G'_{\parallel}}(\mathbf q, \omega, z, z')\nonumber\\
+\sum_{G_{\parallel 1}, G_{\parallel 2}} \int_{-L/2}^{L/2} dz_1 dz_2 \Pi^0_{G_\parallel, G_{\parallel 1}}(\mathbf q, \omega, z, z_1) \nonumber\\
\times  \mathbf D^0_{G_{\parallel 1}, G_{\parallel 2}}(\mathbf q, \omega, z_1, z_2) \Pi_{G_{\parallel 2}, G'_{\parallel}}(\mathbf q, \omega, z_2, z')\nonumber\\
\label{current0}
\end{eqnarray}
where $\mathbf q$ is the momentum transfer vector parallel to the $x-y$ plane and $\mathbf G_\parallel = (G_x, G_y)$ are 2D reciprocal lattice vectors.
Also, the Fourier transformed free-photon propagator is given by \cite{Despoja2009}

\begin{eqnarray}
\mathbf D^0_{G_{\parallel}, G'_{\parallel }}(\mathbf q,  \omega, z, z')= \mathbf D^0 (\mathbf q + \mathbf G_\parallel, \omega, z, z') \delta_{G_{\parallel}, G'_{\parallel }}
\end{eqnarray}
where
\begin{eqnarray}
\mathbf D^0 (\mathbf  q, \omega, z, z') = - \frac{4\pi c}{\omega ^2} \delta(z-z')\mathbf z~\cdot~\mathbf z' \nonumber\\
+ \frac {2\pi i}{c\beta} \{\mathbf e_s~.~\mathbf e_s + \mathbf e_p~.~\mathbf e_p \} e^{i\beta |z-z'|}
\end{eqnarray}

The directions of $\mathbf s$(TE) and $\mathbf p$(TM) polarized fields describe by  $\mathbf e_s = \mathbf q_0 \times \mathbf z$ and $\mathbf e_p = \frac{c}{\omega}[-\beta sgn(z-z')\mathbf q_0 +q \mathbf z]$, where $c$ is the velocity of light, $\beta=\sqrt{\omega^2/c^2-|{\bf q}|^2}$ and the $\mathbf q_0$ is the unit vector in the $\mathbf q$ direction.

Since the integration in Eq. \ref{current0} is performed for $-L/2 < z < L/2$, the current fluctuation created in the region $-L/2 < z < L/2$ can interact via photon propagator $D^0(\mathbf q, \omega, z_1, z_2)$ only with the current fluctuation within the region $-L/2 < z < L/2$, even though the induced electromagnetic field spread over the whole space.  This restriction guarantees that $\Pi_{\mu\nu}$ contains information about only the electromagnetic modes characteristic of the electronic system limited to the region $-L/2 < z < L/2$.
\par
The current-current response tensor $\Pi_{\mu\nu}$ can be obtained by solving the matrix Dyson equation that it is Fourier transformed by Eq. \ref{current0} in the $z$ direction

\begin{eqnarray}
\Pi_{G_z, G'_z}(\mathbf q, \omega)=\Pi^0_{G_z, G'_z}(\mathbf q, \omega)+\sum_{G_{z1}, G_{z2}}\Pi^0_{G_z, G_{z1}}(\mathbf q, \omega)\nonumber\\
\times \mathbf D^0_{G_{z1}, G_{z2}}(\mathbf q, \omega) \Pi_{G_{z2}, G'_z}(\mathbf q, \omega)
\label{current}\nonumber\\
\end{eqnarray}
where
\begin{eqnarray}
\mathbf D^0_{G_z, G'_z}(\mathbf q, \omega)= \frac{1}{L} \int_{-L/2}^{L/2} dz dz' e^{-iG_z z} \mathbf D^0(\mathbf q, \omega, z, z') e^{iG'_z z'}\nonumber\\
\end{eqnarray}

The noninteracting current-current response tensor can also be written as
\begin{eqnarray}
\Pi^0_{\mu\nu, G_z, G'_z}(\mathbf q, \omega)= \frac{1}{\Omega} \sum_{\mathbf k, n, m} \frac{\hbar \omega}{E_n(\mathbf k)-E_m(\mathbf k+\mathbf q)}\nonumber\\
\times [J^\nu_{\mathbf k n , \mathbf k+\mathbf q m} (G'_z)]^* J^\mu_{\mathbf k n , \mathbf k+\mathbf q m} (G_z)\nonumber\\
\times\frac{f_n(\mathbf k)-f_m(\mathbf k+\mathbf q)}{\hbar\omega + i\eta+ E_n(\mathbf k) - E_m(\mathbf k+\mathbf q)}\nonumber\\
\label{current}
\end{eqnarray}
where the current vertices are
\begin{eqnarray}
J^\mu_{\mathbf k n , \mathbf k+\mathbf q m} (G_z)= \int_\Omega d\mathbf r e^{-i\mathbf q . \rho -iG_z z} J^\mu_{\mathbf k n, \mathbf k+\mathbf q m} (\mathbf r)
\end{eqnarray}
and
\begin{equation}
 J^\mu_{\mathbf k n, \mathbf k+\mathbf q m} (\mathbf r) = \frac{\hbar e}{2 im}  \{\phi^*_{n\mathbf k} (\mathbf r) \partial_\mu \phi_{m\mathbf k+\mathbf q} (\mathbf r)
- \partial_\mu {\phi^*_{n\mathbf k} (\mathbf r) \phi_{m\mathbf k+\mathbf q} (\mathbf r)}\}\nonumber
\end{equation}
$\Omega = \mathcal{S} \times L$ is the normalized volume and $f_n(\mathbf k) $ is the Fermi-Dirac distribution at temperature $T$. Notice that we define the three-dimensional
vector ${\bf r}=({\bf \rho}, z)$ and the wave function $\phi_{n\mathbf k}$ is expanded over the plane waves with coefficients which are obtained by solving
the Kohn-Sham equations self-consistently.

We can use a useful relation between the optical absorption and current-current response tensor that was already thoroughly discussed in \cite{Rukelj}.
Now, we want to exploit the normalized absorption power per unit area which is given by

\begin{equation}
\mathcal{A}_{s,p} =  \frac{4\pi}{\omega} S_{s,p}(\mathbf q,\omega),
\label{absor}
\end{equation}
where the dynamical spectral function is
\begin{eqnarray}
S_{s,p}(\mathbf q,\omega) =\Im m \left\{\sum_{\mu,\nu} e^{s,p}_{\mu} e^{s,p}_{\nu} \sum_{G_z,G'_z} I^+_{G_z} \Pi_{\mu,\nu, G_zG'_z}(\mathbf q,\omega) I^+_{G'_z}\right\}\nonumber\\
\label{absor1}
\end{eqnarray}
in which, the form factors are
\begin{equation}
I^\pm_{G_z}= \frac{2}{\sqrt L} \frac {\sin[(\beta\pm G_z)L/2]}{\beta \pm G_z}
\label{absor2}
\end{equation}

By using the energy conservation law, the transmitted electromagnetic energy flux can be calculated by
\begin{equation}
\mathcal{T}_{s,p}= 1 - \mathcal{R}_{s,p} - \mathcal{A}_{s,p}
\end{equation}
where $\mathcal{R}_{s,p}$ is the reflected energy flux and given by
\begin{equation}
\mathcal{R}_{s,p} = \mid R_{s,p} \mid^2
\end{equation}

Here, the amplitude of the reflected $s$ and $p$ waves are
\begin{eqnarray}
&&R_s = \frac{2\pi i}{c\beta} D_{xx}(\mathbf q, \omega),\nonumber\\
&&R_p = \frac{2\pi i}{\omega}[D_{yy}(\mathbf q, \omega) \cos\theta - D_{zz}(\mathbf q, \omega) \sin\theta \tan\theta],
\nonumber\\
\end{eqnarray}
where $\theta=\sin^{-1}(\beta/q)$ is the angle between wave vector ${\bf q}$ and the normal vector of the system surface.  The surface electromagnetic field propagator is defined by
\begin{equation}
D_{\mu\nu}(\mathbf q, \omega) = \sum_{G_zG'_z} I^+_{G_z} \Pi_{\mu\nu, G_z, G'_z}(\mathbf q, \omega) I^-_{G'_z}
\end{equation}

At the end of this section, it is worth mentioning that the dielectric tensor and the conductivity tensor can be obtained with the current-current response. The first one is defined as~\cite{Despoja2011}
\begin{equation}
\varepsilon_{\mu\nu}(\omega)= \delta_{\mu\nu}+\frac{4\pi }{\omega^2}~ \Pi^0_{\mu\nu,00}(\mathbf q=0, \omega)\label{eps}
\end{equation}
which describes the response of phosphorene to an external homogeneous electrical field. With knowledge of the optical dielectric function, the refractive index of the system can be obtained. Moreover, the real part of the dielectric function describes the imaginary part of the optical conductivity, however, the imaginary part of
the dielectric function describes the absorption of light.

Since $\varepsilon_{\mu\nu}(\omega)= \delta_{\mu\nu}+(4\pi/\omega)i\sigma(\omega)$, the longitudinal conductivity is given by
\begin{equation}
\Re e \sigma_{\mu\mu}(\omega)=\frac{1}{\omega}\Im m \Pi_{\mu\mu,00}(\mathbf q=0 , \omega)\label{opc1}
\end{equation}
which corresponds to the experimentally measurable optical conductivity~\cite{Novko2016}. The optical longitudinal conductivity includes the interband transitions and the contribution of the intraband
transitions, which leads to the fact that the Drude-like term is no longer relevant in this study since the momentum relaxation time is assumed to be infinite. This approximation is valid at low temperature and for a clean sample where defect, impurity, and phonon scattering mechanisms are ignorable.

\section{NUMERICAL RESULTS AND DISCUSSIONS}\label{sec:result}

In this section, we present our numerical results for few-layer black phosphorus, based on first principles simulations within the DFT-RPA approach at zero temperature.
The band structure of few-layer phosphorene is plotted in Fig. \ref{fig2}. Clearly, the band characteristic of the AB bilayer and ABA trilayer phosphorene are similar to that of the monolayer one, except that in the bilayer and trilayer, energy level splitting occurs due to the interlayer interactions. Our DFT calculations predict that the band gap of monolayer phosphorene is direct and equal to 0.98 eV and intriguingly, it decreases to 0.6 and 0.52 eV for bilayer and trilayer phosphorene, respectively. The nature of the band gap remains direct for trilayer to single-layer black phosphorous irrespective of the nature of stacking and it is more promising in terms of applying tunnel field-effect transistor devices~\cite{Sengupta}.

In the 2D case, the isofrequency profiles are obtained by horizontally cutting the dispersion surface. Also, the isofrequency contour plots of the energies around the conduction band minimum (CBM) and the valence band maximum (VBM) in the $k-$space of few-layer phosphorene are illustrated in Figs. \ref{fig2}(d) - \ref{fig2}(f). Most importantly, the shape of the Fermi surfaces in the systems is almost an elliptic shape, especially at low charge density. The elliptic shape of the Fermi surfaces, particularly in the hole-doped case demonstrates the anisotropic band energy dispersion of the few-layer phosphorene along the $\Gamma X$ and $\Gamma Y$ directions. It turns out that, in the hole-doped cases, the anisotropic band energy dispersion decreases by increasing the number of the layers.
\par
It is well-known that DFT underestimates the band gap of semiconductors and the advance Green's function method can provide improved predictions. The Green's function gap of monolayer phosphorene calculated by Tran {\it et al}.~\cite{Tran2014} is about 2.0 eV. A recent Monte Carlo study of monolayer phosphorene \cite{Frank}, using infinite periodic superlattices as well as finite clusters, predicted that the band gap is 2.4 eV.

\subsection{Optical absorption and transmission of bilayer phosphorene}
We explore the optical absorption calculated by using Eqs.~\ref{absor}-\ref{absor2} with the current-current response tensor $\Pi_{\mu\nu}$ (Eq.~\ref{current}) .
Having calculated the structure of few-layer phosphorene, we can obtain the Kohn-Sham wave functions $\phi_{m\mathbf k} (\mathbf r)$ and energies $E_m (\mathbf k)$ which are invoked to calculate the current-current response, $\Pi_{\mu\nu}$.
In the summation over $\mathbf k$ in Eq. \ref{current}, we use $120\times 160\times 1$ $k$-point mesh sampling and band summation ($m$, $n$ ) is performed over $30$ and $40$ bands for bilayer and trilayer phosphorene, respectively. The damping parameter, $\eta$ is $50$ meV in all figures, unless we specifically define this value otherwise.

\begin{figure}
\includegraphics*[width=7.5cm]{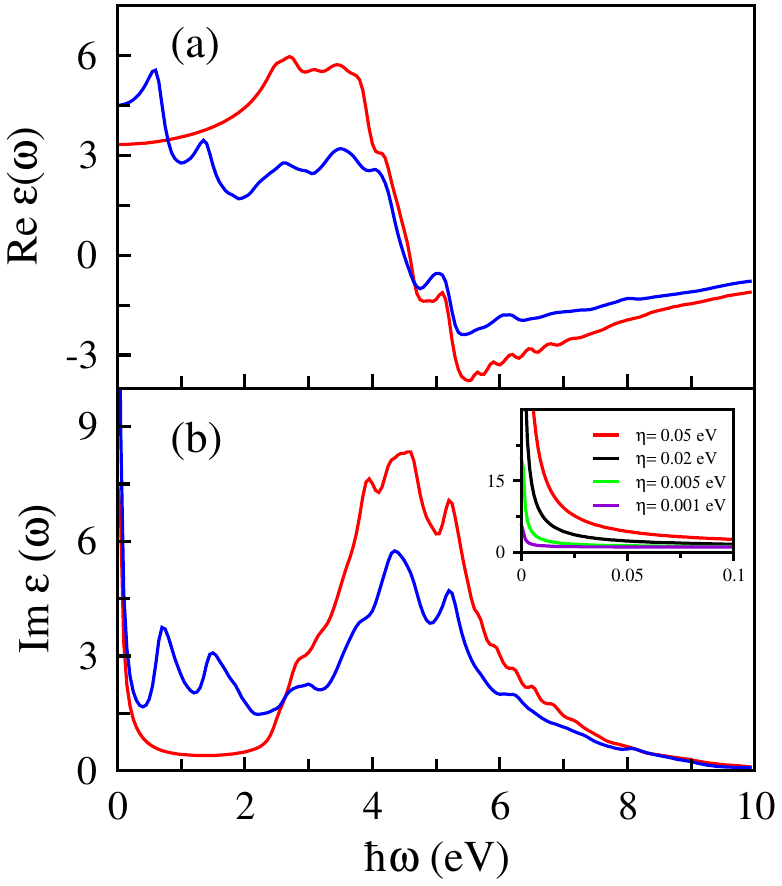}
\caption{(Color online)  (a) The real and (b) imaginary  parts of the  dielectric tensor of bilayer phosphorene vs $\omega$ for $\varepsilon_{xx}$ (blue line) and $\varepsilon_{yy}$ (red line). Note that the interband transitions are included in the dielectric tensor. The imaginary part of the dielectric function illustrates the excitations in the system. Note that, the peaks in the imaginary part of the dielectric functions correspond to the drops in the real part of the dielectric functions. The inset shows  $\Im m \varepsilon_{yy} (\omega)$ around $\omega=0$ for different values of $\eta$. The $\Im m \varepsilon_{yy} (\omega=0)$ tends to zero by decreasing the value of $\eta$.
	\label{fig3}}
\end{figure}

The real and imaginary parts of the {\it ab initio} DFT dielectric function (Eq. \ref{eps}) are shown in Fig \ref{fig3}. The dielectric function is calculated by invoking the Kohn-Sham wave functions and they change by changing the structures and inter-atomic interactions. Therefore, the
dielectric function depends on the material density and the
interlayer distance.

 We ought to note that at the onset of transparency at the plasmon frequency we have $\Re e \varepsilon_{xx/yy}(\omega)=0$. The imaginary part of the DFT dielectric function shows the excitations in the system. The $\Im m \varepsilon_{xx}(\omega)$ displays considerable structure before decreasing to become nearly zero at $\hbar \omega \approx 2$ eV. This leads us to imagine that below 2 eV, $\Im m \varepsilon_{xx}(\omega)$ is due to the topmost occupied electron levels. Electrons above 10 eV play no significant part in the optical spectrum.
As shown in Fig. 3, the peaks in the imaginary part of the dielectric function are located in the energy range between  $4$ and $6$ eV, where the absorptions are maximum, which corresponds to the drop of the real part of the dielectric functions. The values of the extremum positions for both imaginary and real parts
of the dielectric function are equal as they are related by
the Kramers-Kroning relations.
It is worth to mention that the finite
values of $\Im m \varepsilon (\omega)$ around $\omega=0$ are a numerical artifact and it should be zero. This  artifact is related to the finite value of $\eta$, and from $1/\omega^2$ factor in Eq. \ref{eps} and it can be reduced by decreasing of $\eta$~\cite{Sangalli}. In the inset of Fig. \ref{fig3}(b), we illustrate the imaginary part of $\varepsilon_{yy}(\omega)$ for different values of $\eta$. The role of $\eta$ in this problem is completely clear.
\par
There is an important $f$-sum rule for the dielectric function which is used in the analysis of the absorption spectral. A general $f$-sum rule for the imaginary part of the dielectric function says that
\begin{equation}
\int_0^{\infty} \omega \Im m\varepsilon_{\alpha \beta} d\omega=\delta_{\alpha \beta}\frac{2N \pi^2 e^2}{V m}
\end{equation}
where $N=20$ is the number of electrons, $V$ is the unit cell volume and $m$ is the electron bare mass. We validate our numerical DFT-RPA results by considering 200 band structures in each system and find that the theory gives satisfactory results by 6\%.

\begin{figure}
\includegraphics*[width=7.5cm]{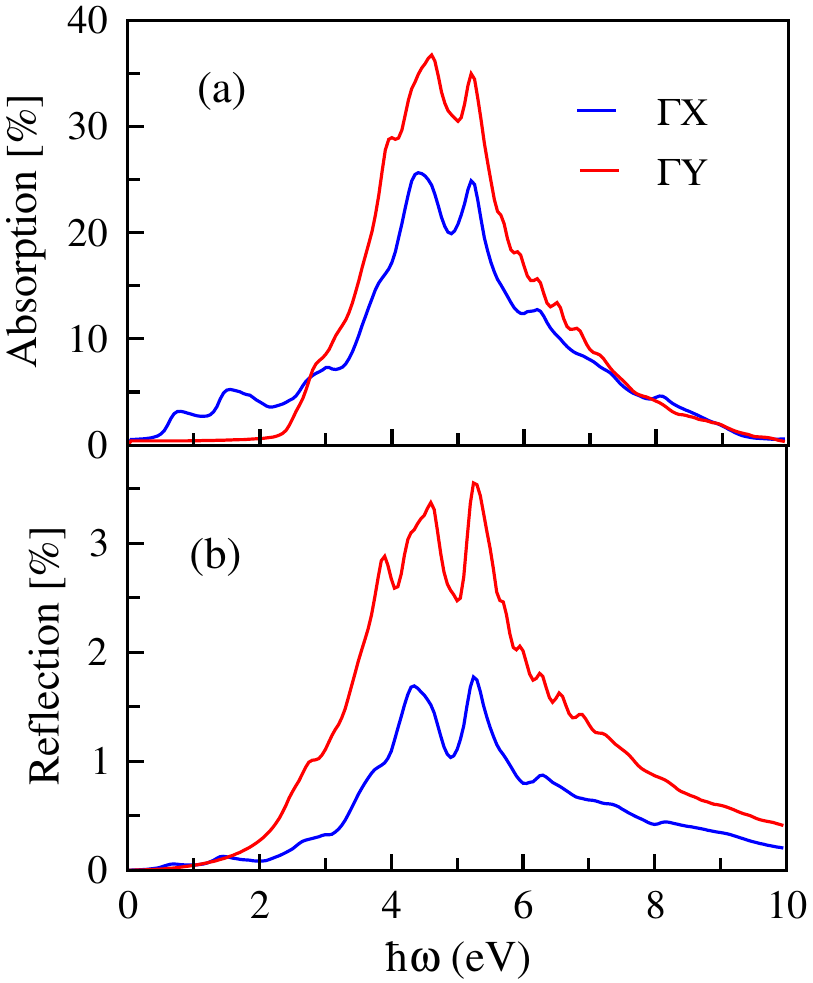}
\caption{(Color online) Optical (a) absorption and (b) reflection of pristine bilayer phosphorene of the $s$-polarized (along the armchair direction) and  $p$-polarized (along the zigzag direction) for normal incident ($\theta=0$). Noticeably, the optical transition of the valence band to the conduction band appears along the armchair direction at about $0.72$ eV, while it is zero up to $2.5$ eV along the zigzag direction. \label{fig4}}
\end{figure}

The optical absorption and reflection of the $p-$ and $s$-polarized normal incidence ($\theta= 0$) are shown in Fig. \ref{fig4} as a function of photon energy ($\hbar \omega$) for bilayer phosphorene. The optical absorption illustrates a maximum around energy between $4-6$ eV where the most profound photon-induced absorption of electrons between the mentioned bands occurs. The optical transition of the valence band to the conduction band appears along the armchair direction at about $0.72$ eV, while it is zero between $0$ and $2.5$ eV along the zigzag direction. This phenomenon originates from the contribution of $\Im m \varepsilon_{xx}(\omega)$ in this region and consistent with experimental observations reported in Ref. \cite{Buscema}. These results demonstrate a strong linear dichroism in bilayer phosphorene, where the position of the lowest energy absorption peaks for the armchair and zigzag directions differ significantly.

It can be found from the symmetry of the wavefunctions that dipole operator connects the VBM and CBM states for the $s$-polarization, allowing the direct-band gap process, however, this is symmetry-forbidden for the $p$-polarization and transition occurs between the VBM and CBM states elsewhere in the Brillouin zone~\cite{qiao}. Therefore, we expect that phosphorene would be a suitable material for applications associated with liquid-crystal displays and optical quantum computers~\cite{knill,zeng}.

It should be noted that the reflection part of the incident light is negligible as shown in Fig. \ref{fig4}(b). Therefore, the main part of the optical properties goes to the transmission and absorption parts.

\begin{figure}
	\includegraphics*[width=7.5cm]{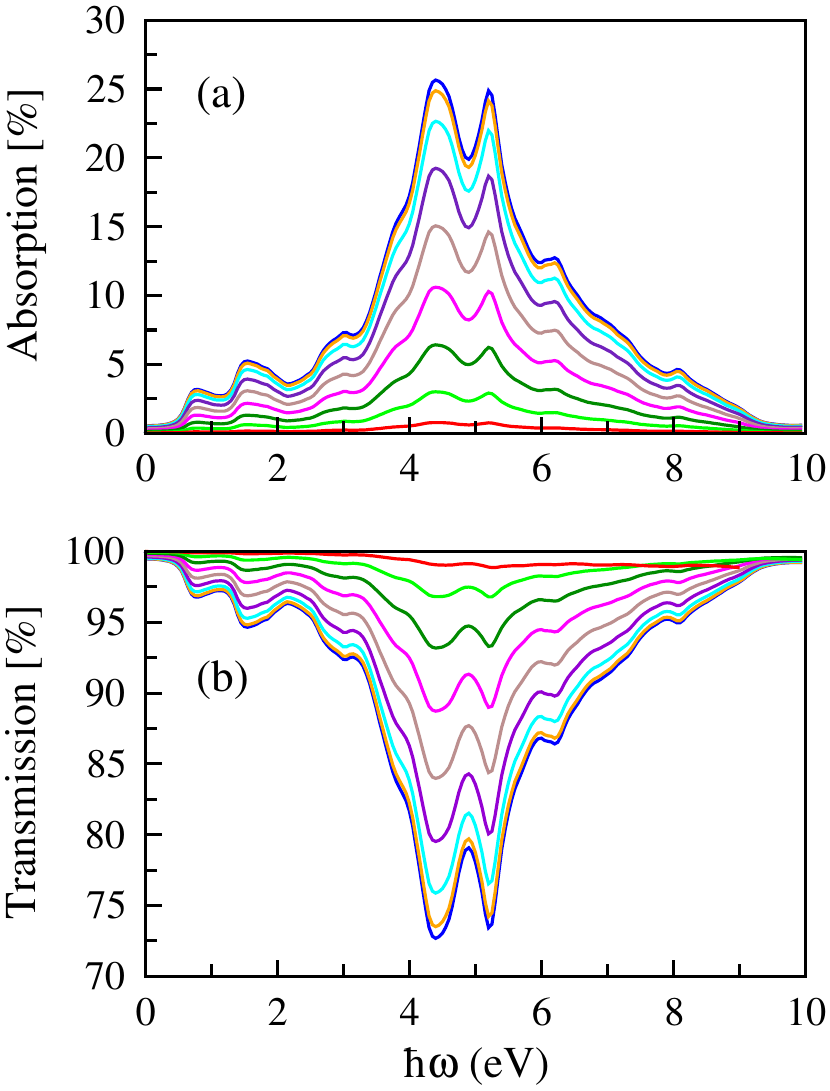}
	\caption{(Color online) Angle-resolved absorption and transmission of bilayer phosphorene with incident angle $\theta =n\Delta\theta$ for $\Delta \theta=10^{\circ}$ and $n=0, 1, ...,8$. The optical absorbance monotonically decreases, as the incident angle of light increases, however, the transmission increases. Notice that phosphorene is transparent when it is illuminated by near grazing incidence light. \label{fig5}}
\end{figure}

We also analyze the optical absorption and transmission of bilayer phosphorene as functions of the incident angle $\theta$ and photon energy $\hbar \omega$ along the armchair direction (Fig. \ref{fig5}). It is observed that the optical absorbance monotonically decreases, as the incident angle of light increases, however, the transmission increases. In particular, Fig. \ref{fig5} shows that bilayer phosphorene is transparent when it is illuminated by a nearly grazing incidence light. It is intriguing that the peaks in the absorption spectra correspond to the dips in the transmitted spectra and the peaks show that the reflected light is negligible. Similar physical behaviors have been obtained for few-layer phsophorene and monolayer MoS$_2$ \cite{Rukelj}.

We would like to recall well-known dipole selection rules, which determines whether transitions are alloweded or forbidden based on the symmetry of the valence or conduction wave functions.
The optical absorption can be calculated in the dipole-transition approximation as~\cite{Tran2015}
\begin{eqnarray}
\alpha(\omega)= \frac{(2\pi)^2}{\hbar\omega} \sum_{\nu c \mathbf k} |<c \mathbf k|\hat{e} \cdot\mathbf p|\nu\mathbf k>|^2 \delta(\hbar\omega-E_{c\mathbf k}+E_{\nu\mathbf k})\label{opc}
\end{eqnarray}
where $\mathbf p$ is the dipole matrix operator and $\hat{e}$ is the direction of the polarization of the incident light. The dipole selection rules allow transitions in which angular momentum between the valence and the conduction states differ from unity. Since the parity of $\mathbf p$ is odd, two wavefunctions of the valence and the conduction have opposite parities in the direction of $\hat{e}$ and the same parity in other directions. It is worth mentioning that in order to have a significant absorption at a particular energy, the joint density of states, $\frac{1}{V}\sum_{\mathbf k \nu c} \delta(\hbar \omega+E_{c\mathbf k}-E_{\nu\mathbf k})$, must have a Van-Hove singularity for a given energy. Most importantly, this one-particle picture of the transition process is totally inadequate and does not come close to describing the absorption spectra observed in experiments. Consequently, our analysis based on the many-body optical absorption is needed.

Note that when states near the VBM or the CBM have multicomponent characters, the spinors describing these components can pick up nonzero winding numbers and in such systems, the strength and required light polarization of an excitonic optical transition are dictated by the optical matrix element winding number. This winding-number physics, which mainly emerge in nanoribbon structures, leads to novel exciton series and optical selection rules~\cite{Cao2018}. In this work, we focus on only the sum-rule selection rules presented in Eq. \ref{opc}.
\par
For bilayer phosphorene, the percent contribution from each atomic orbital to the valence and the conduction wavefunctions at the special $k$-point listed in Tabel. \ref{tab:table1}. The optical absorption edge along the armchair direction is related to an interband transition from the VBM to the CBM and the allowed transitions are $d_{x^2-y^2}\rightarrow p_x$, $d_{z^2}\rightarrow p_x$ and  $d_{zx}\rightarrow p_z$.
\par
For a zigzag polarization, on the other hand, the optical absorption edge around $2.5$ eV is from the VBM to CBM+1 (the next band higher in energy than the conduction band) at a $k$ point along the $\Gamma-Y$ direction which mainly originates from the interband transition. In this case, CBM+1 contains components of $p_y$ and $d_{xy}$ which it causes allowed transitions such as $p_y \rightarrow s$, $p_y \rightarrow d_{x^2-y^2}$ and $p_z \rightarrow d_{xz}$.

A similar analysis of the optical transition is applicable for the trilayer phosphorene based on the details given in Table I. \ref{tab:table1}.

\begin{widetext}
\begin{table*}[ht]
\caption{\label{tab:table1} The PDOS of the wavefunctions that contribute to allowed transitions along the armchair and zigzag directions. The percentage contributions from atomic orbitals of the wavefunctions are determined.}
\begin{center}
\begin{tabular}{cccccccccccccccc}
\hline\hline
 &type& direction & $\mathbf k$(X,Y) &~~ state~~ &~~ $s$~~ &~~ $p_z$~~ & ~~ $p_x$ ~~& ~~$p_y$~~ &~~ $d_{z^2}$~~   & ~~ $d_{xz}$ ~~  & ~~$d_{yz}$ ~~   & ~~$d_{x^2-y^2}$ ~~   & ~~  $d_{xy}$~~ \\  \hline
&bilayer& armchair   &~~(0,0)& CBM      &  6 &  42    &  17   &    0   &   28      &     4        &       0     &     1      &    0      \\
   &&         &        &  VBM    &  8 &  72    &   4   &    0   &   0       &     10       &       0     &     5      &    0       \\ \hline
&bilayer& zigzag    &~~(0,0.35)& CBM+1  &  2 &  24    &   6   &    21  &   1       &     7        &       3     &     6      &    26      \\
    &&             &        & VBM     &  12&  70    &   1   &    10   &   0       &     1        &       0     &     1      &    5       \\
\hline\hline
&trilayer& armchair   &~~(0,0)& CBM      &  9 &  38    &  16   &    0   &   30      &     6        &       0     &     1      &    0      \\
&   &&     &  VBM    &  7 &  77    &   3   &    0   &   0       &     5       &       0     &     5      &    0       \\ \hline
&trilayer& zigzag    &~~(0,0.38)& CBM+1  &  5 &  7    &   19   &    24  &   3       &     6        &       3     &     6      &    26      \\
&   &&     & VBM     &  6&  72    &   1   &    0   &   0       &     1        &      8     &     1      &    8       \\
\hline\hline
\end{tabular}
\end{center}
\end{table*}
\end{widetext}

\par
\begin{figure}
	\includegraphics*[width=7.5cm]{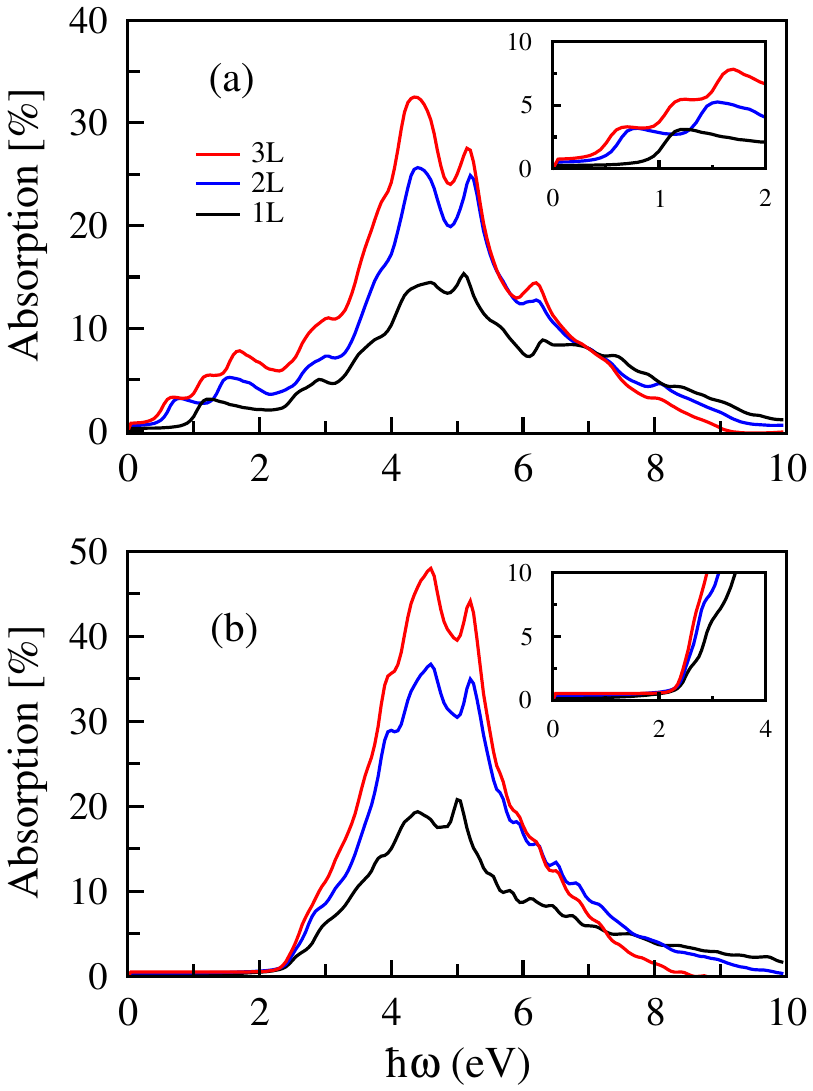}
	\caption{(Color online)Optical absorption spectra of few-layer phosphorene along the (a) armchair and (b) zigzag directions. Light absorptivity improves by increasing the number of the layers in few-layer phosphorene. The inset shows the low energy part of optical absorption. It shows that the absorption spectrum is red-shifted with increasing the number of layers along the armchair direction, while it changes slightly with the addition of phosphorene layers along the zigzag direction.\label{fig6}}
\end{figure}

The optical absorption for the normal incidence ($\theta= 0$) of light of a few-layer phosphorene along the armchair and zigzag directions are compared in Fig. \ref{fig6}. It can be noticed that the optical absorption spectra of the bilayer and trilayer are generally similar to that of the monolayer phosphorene. It is also noticeable that there is a general trend of an increase in the absorption by increasing the number of layers. This result shows that light absorptivity can be improved by appropriately increasing the number of layers in few-layer phosphorene~\cite{Sengupta}.
In the low-energy zone, the absorption spectrum is red-shifted by increasing the number of the layers along the armchair direction, while it changes slightly with the addition of phosphorene's layers along the zigzag direction.

This phenomenon originates from the decreasing of the energy band gap with increasing the number of the layers.
The optical absorption edges are found to start approximately from the band gap. Therefore, the optical absorption edge can be tuned by changing of layers in few-layer phosphoren.

\begin{figure}
\includegraphics*[width=7.5cm]{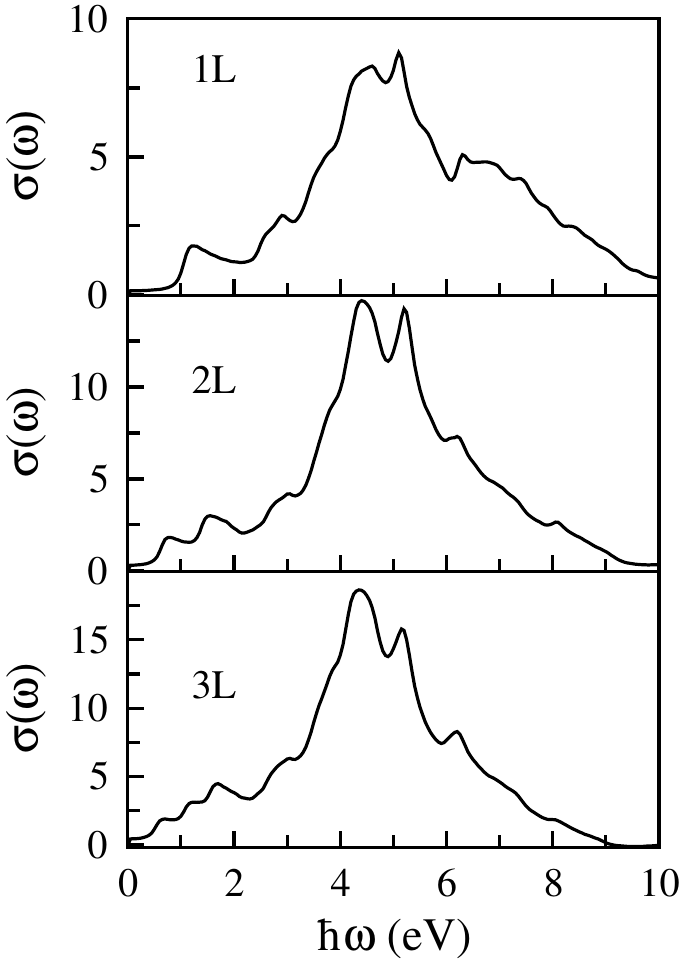}
\caption{(Color online) The real part of the optical conductivity of few-layer phosphorene, in units of $e^2/4\hbar$, along the armchair direction. Notice that since $\Pi^0(q=0,\omega)=\Pi(q=0,\omega)$, the absorption becomes proportional to the optical conductivity, in phosphorene structures.\label{fig7}}
\end{figure}

\begin{figure}
	\includegraphics*[width=7.8cm]{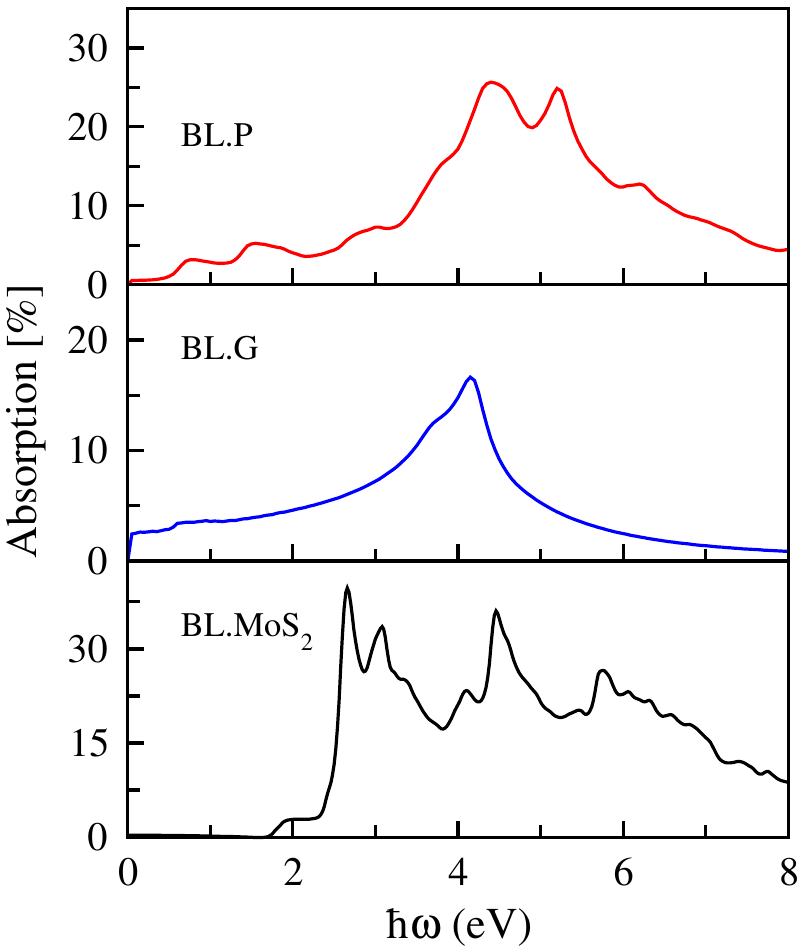}
	\caption{(Color online) The optical absorption of pristine bilayer phosphorene, graphene and molybdenum disulfide along armchair direction for normal incidence ($\theta=0$). Here, $\eta=30$ meV and the value of the peak of the optical absorption depends on the damping constant $\eta$ used in the calculation. It would be noticed that the bilayer phosphorene exhibits greater absorbance compared to the optical absorption of bilayer graphene in the ultraviolet region.}\label{fig8}
\end{figure}

\begin{figure}
	\includegraphics*[width=8.8cm]{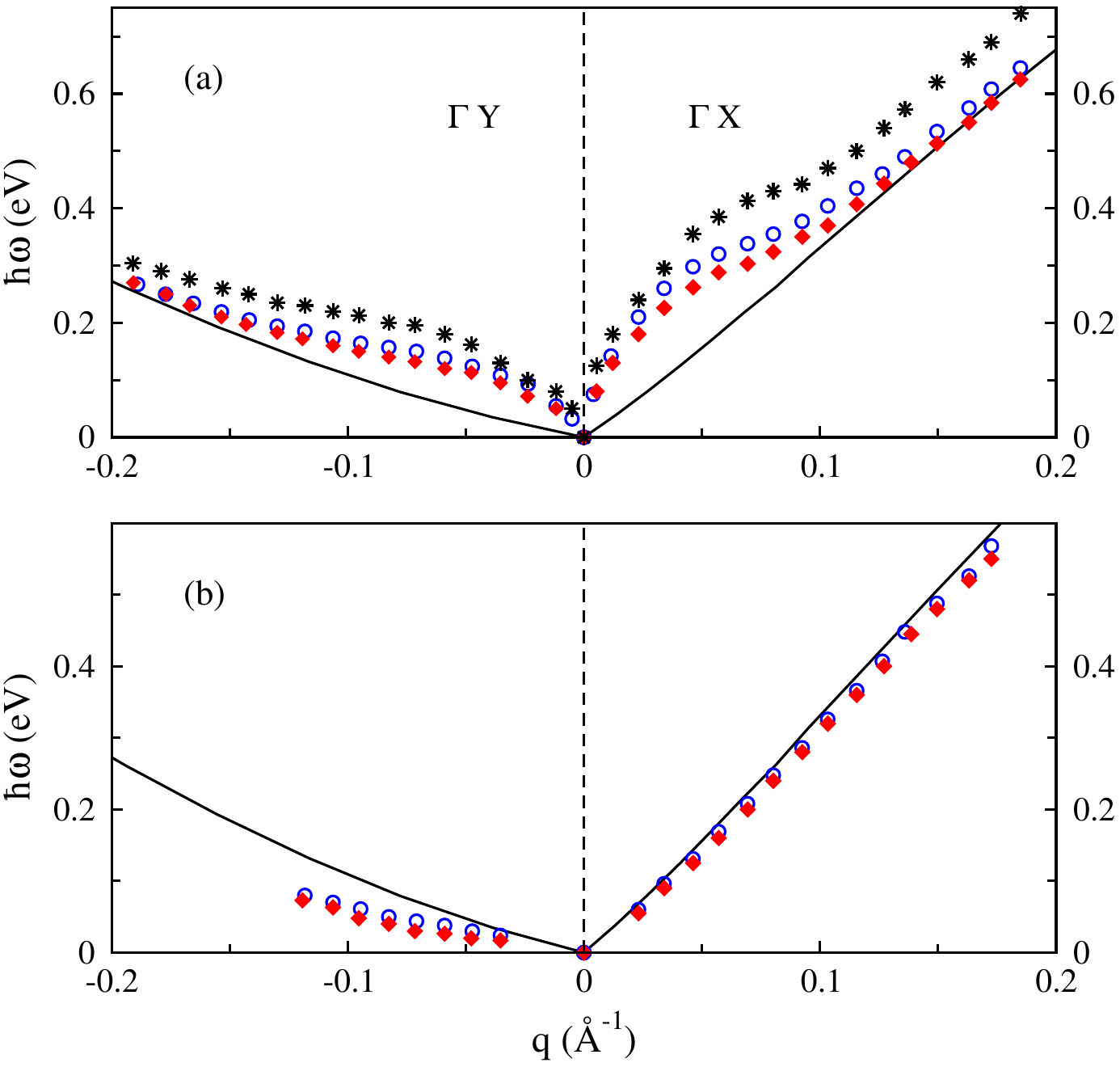}
	\caption{(Color online) (a) The optical and (b) acoustic plasmon modes of few-layer phosphorene for $E_{\rm F}= 0.09$ eV. The diamond symbols, blue circle and star symbols refer to plasmon modes of trilayer, bilayer and monolayer phosphorene respectively. Owing to the anisotropic band structure of few-layer phosphorene along the zigzag and armchair directions, we have an anisotropic plasmon mode in two directions. Notice that $\omega_p$ (monolayer) $\ge$ $\omega_p$ (bilayer) $\ge$ $\omega_p$(trilayer) for optical plasmon modes.}\label{fig9}
\end{figure}

The optical conductivity of few-layer phosphorene is calculated by using Eq.~\ref{opc1}. It is illustrated in Fig. \ref{fig7} for $q= 0$, the homogeneous electrical field directed in the $x$ direction. It is clear there is no new feature regarding the optical conductivity in comparison with the absorption. As mentioned in Ref.~\cite{Novko2016}, the unscreened ($\Pi^0$) becomes equal to the screened ($\Pi$) current-current response  for $q= 0$, and $\Pi^0$ is nonzero only for $G_z =G_{z'}= 0$ where the form factor eventually becomes a constant. Therefore, it can be concluded that the absorption becomes proportional to the optical conductivity.
\par
In Fig. \ref{fig8}, we compare the optical absorption of bilayer phosphorene with those of graphene and molybdenum disulfide for $\eta=30$ meV. As shown in the case of graphene, the absorption onset starts from $0$ eV which is due to the gapless dipole active $\pi$ to $\pi^*$ interband transitions near the $K$ point of the Brillouin zone. However, it is nearly zero in the region between $0-0.72$ eV for bilayer phosphorene and $0-1.86$ eV for bilayer MoS$_2$. In fact, it shows semimetal nature in bilayer graphene and semiconductor characteristic of bilayer phosphorene and bilayer MoS$_2$.

Regarding bilayer graphene, in the infrared region, the spectral absorption per pristine graphene layer is a constant, $\pi\alpha=2.3\%$ ($\alpha$ is the fine-structure constant), in good agreement with that obtained in a recent experiment and theoretical predictions~\cite{Yang,Nair}. Obviously this value is valid for perfect graphene flake and depends strongly on the damping constant $\eta$ used in the calculation.
In the visible energy, the absorption monotonically increases. The first absorption maximum, which appears in the ultraviolet region at $\omega = 4.17$ eV, is a consequence of the dipole active interband  $\pi$ to $\pi^*$ transitions along the MM' and M$\Gamma$ directions of the first Brillouin zone, as discussed in details in Ref~\cite{Novko_2015}.

Bilayer phosphorene exhibits greater absorbance compared to the optical absorption of AB bilayer graphene in the ultraviolet region. The same results pertain to their monolayer attribute \cite{Lin}. Therefore, these results reveal that phosphorene absorb light strongly and it is a promising material to utilize in thin-film solar cells and photoelectric converters~\cite{Dai_bi}.
\par
The onset of optical absorption in AB bilayer MoS$_2$ is about $1.86$ eV and slowly increasing to a plateau, which corresponds to a transition of the valence band to the conduction band around the $K$ point. The intense peak is at 2.7 eV which is red-shifted compared to the intense peak in the monolayer. It is worth mentioning that the maximal peak of the absorption of bilayer MoS$_2$ is in the visible region, while bilayer graphene and phosphorene are transparent in this region.

Finally, we use the formalism that was presented in previous works \cite {Novko_2015,torbatian,torbatian2018} and calculate the collective modes of few-layer phosphorene. To do so, the Fermi energy is set to be $0.09$ eV above the edge of the conduction band of each phosphorene structure and our numerical results are illustrated in Fig.~\ref{fig9}.
One mode is  the  optical plasmon mode, which has its counterpart
in single-layer samples with $\omega(q\rightarrow 0)~\sqrt{q}$.
This mode corresponds
to a collective excitation of the electron gas in which the
carriers of both layers oscillate in-phase. The optical plasmon modes increase by doping the system.
Although both optical plasmon modes along the $\Gamma X$ and $\Gamma Y$ directions follow a low-energy $\sqrt q$ dependence~\cite{Low}, the plasmon in the armchair direction has a higher density than that the zigzag and armchair directions make the anisotropic plasmon mode in both two directions \cite{Rodin}. Also, it can be seen that the optical plasmon mode of $\omega_p$ (monolayer) $\ge$ $\omega_p$ (bilayer) $\ge$ $\omega_p$(trilayer). It is related to the electron concentration that those systems contain. Basically, with $E_{\rm F}= 0.09$ eV, the electron concentration of monolayer phosphorene is greater than that in the bilayer and thus the trilayer has the lowest electron concentration at given the Fermi energy. In addition to the optical mode, we observe the existence of an additional mode in the excitation spectrum dispersion for bilayer and trilayer phosphorene. In Fig.~\ref{fig9}(b), the acoustic modes of bilayer and trilayer phosphorene along the $\Gamma X$ and $\Gamma Y$ directions are shown. These modes correspond to a collective oscillation in which
the carrier density in the two layers oscillates out-of-phase. These modes with a low-energy nearly linear dispersion are damped as they lie on the electron-hole continuum region.

\section{CONCLUSION}\label{sec:concl}
In this work, we have analyzed the angle-resolved optical absorption and transmission of few-layer phosphorene using the current-current response tensor calculated in the framework of {\it ab initio} DFT calculations and the many-body random phase approximation. The optical transition of the valence band to the conduction band appears along the armchair direction at about 0.72 eV, while it is zero between 0 and 2.5 eV along the zigzag direction in bilayer phosphorene.
In few-layer phosphorene, it is observed that the optical absorbance monotonically decreases, as the incident angle of light increases, and is transparent when it is illuminated by near grazing incidence of light. But the transmission increases,  as the incident angle of light increases and it becomes almost $100\%$ for near grazing incidence. It can be noticed that the optical absorption spectra increases by increasing the number of layers. In the low-energy zone, the absorption spectrum is red-shifted by increasing the number of layers along the armchair direction, while it changes slightly along the zigzag direction. Also, we have compared the optical absorption of bilayer phosphorene with those of bilayer graphene and MoS$_2$. It is shown that the bilayer phosphorene exhibits greater absorbance than the optical absorption of bilayer graphene in the ultraviolet region. The maximal peak in the absorption of bilayer MoS$_2$ is in the visible region, while bilayer graphene and phosphorene are transparent in this region. Moreover, the anisotropy of the band structure of few-layer phosphorene along the armchair and zigzag directions is manifested in the collective plasmon excitations. Our results provide a microscopic understanding of the electronic and optical characteristics of few-layer phosphorene.

\section{acknowledgments}
We thank H. Akbarzadeh for fruitful discussions and his help. Z. T. would like to thank the Iran National Science Foundation for its support. This work is also supported by the Iran Science Elites Federation.


\end{document}